\documentclass[12pt]{article}
\usepackage{amsfonts,amssymb}

\def\C{\mathbb{C}}
\def\N{\mathbb{N}}
\def\Z{\mathbb{Z}}

\def\g{\mathfrak g}

\def\bq{ \begin{equation} }
\def\eq{ \end{equation} }
\def\ben{ \begin{eqnarray} }
\def\en{ \end{eqnarray} }

\def\frac#1#2{{#1\over #2}}

\def\on#1#2{\mathop{\vbox{\ialign{##\crcr\noalign{\kern2pt}
$\scriptstyle{#2}$\crcr\noalign{\kern2pt\nointerlineskip}
\kern-2pt$\hfil\displaystyle{#1}\hfil$\crcr}}}\limits}

\begin{document}

\baselineskip=15pt
\vspace{1cm} \centerline{{\LARGE \textbf {Systems of Gibbons-Tsarev type
 }}}
\vspace{0.4cm}
  \centerline{{\LARGE \textbf {and
integrable 3-dimensional models
 }}}

\vskip1cm \hfill
\begin{minipage}{13.5cm}
\baselineskip=15pt {\bf A.V. Odesskii ${}^{1,}{}^{2}$,  V.V. Sokolov ${}^{1}$}
\\ [2ex] 
{\footnotesize
${}^{1}$  L.D. Landau Institute for Theoretical Physics (Russia)
\\
${}^{2}$ Brock University (Canada)
\\}
\vskip1cm{\bf Abstract}
We review the role of Gibbons-Tsarev-type systems in classification of integrable multi-dimensional hydrodynamic-type systems. Our main observation is an universality of Gibbons-Tsarev-type systems. We also constract explicitly a wide class of 3-dimensional hydrodynamic-type systems corresponding to the simplest possible Gibbons-Tsarev-type system.

\end{minipage}

\vglue1cm \textbf{Address}: L.D. Landau Institute for Theoretical
Physics of Russian Academy of Sciences, Kosygina 2, 119334,
Moscow, Russia

\textbf{E-mail}: aodesski@brocku.ca, sokolov@itp.ac.ru
\newpage

\section{Introduction}

Integrable equations play important role in both Mathematics and Physics. Unfortunately, rigorous
and universal definition of integrability applicable in all situations does not exist. Different viewpoints on the integrability 
can be found in \cite{what,int}.
It is well known that integrable 2-dimensional PDEs
\begin{equation}   \label{PDE}
u_t=F(u, u_{x}, ..., u_{nx} )
\end{equation}
like the KdV-equation, for any $N$ possess families of exact solutions
depending on arbitrary constants $c_{1},...,c_{N}$. All these finite-gap and
solitonic-type
solutions can be constructed by so called ODE-reductions. A pair of
compatible $N$-component systems of ODEs
\begin{equation}   \label{ODEred}
r_{x}^{i}=f^{i}(r^{1},..., r^{N}), \qquad r_{t}^{i}=g^{i}(r^{1},...,
r^{N}), \qquad i=1,..., N
\end{equation}
is called an ODE-reduction of (\ref{PDE}) if there exists a function
$U(r^{1},...,r^{N})$ such that $u=U(r^{1}(x,t),...,r^{N}(x,t))$ satisfies
(\ref{PDE}) for any solution  $r^{1}(x,t),...,r^{N}(x,t)$ of
(\ref{ODEred}). It is clear that the solution $u$ depends on $N$ arbitrary
parameters being initial values for (\ref{ODEred}) at generic point. The
existence of special ODE-reductions for arbitrary $N$ can be chosen as a criterion
of integrability for equation (\ref{PDE}). For example, one can assume that (\ref{PDE}) admits a series of
differential ODE-constraints
$$
u_{m,x}=G_m(u,u_x,...,u_{m-1,x}),
$$
where $m$ is arbitrary. Clearly, equation (\ref{PDE}) and the ODE-constraint can be rewritten as a pair of compatible dynamical systems with respect to $x$ and $t$. Another example of ODE-reductions is provided by Dubrovin's equations \cite{dub1}. However, in the 2-dimensional
case there exist more efficient and constructive integrability criteria, like the existence
of higher local symmetries or conservation laws (see \cite{miksok} and references therein).

If the number $d$ of independent variables is greater then 2, then higher {\bf local}
symmetries for integrable models 
do not exist (for some generalization of the symmetry approach
to the case of non-local symmetries see \cite{mikh}). In such a
situation the existence of $N$-component reductions can be regarded as one of the
most powerful methods of searching for new integrable models. Notice that
 one has to consider for the reductions some compatible systems of
PDEs of dimension $\le d-1$ instead of ODEs (\ref{ODEred}).

In \cite{ferhus1} this approach has been systematically applied to
some classes of 3-dimensional systems of the form
\begin{equation}   \label{genern}
\sum_{j=1}^n a_{ij}({\bf u})\,u_{j,t}+\sum_{j=1}^n b_{ij}({\bf
u})\,u_{j, y}+ \sum_{j=1}^n c_{ij}({\bf u})\,u_{j, x}=0,
\qquad i=1,...,n+k,
\end{equation}
where ${\bf u}=(u_{1},\dots, u_{n}),$ and  $k \ge 0.$ Pairs of compatible diagonal semi-Hamiltonian (see formula (2.12))
hydrodynamic-type systems of the form
\begin{equation}\label{gidra}
r_{t}^{i}=v^{i}(r^{1},..., r^{N})r_{x}^{i} \qquad r_{y}^{i}=w^{i}(r^{1},..., r^{N})r_{x}^{i} \qquad i=1,2,...,N,
\end{equation}
have been taken for reductions. According to definition of reductions, the
corresponding solutions of (\ref{genern}) are determined by some
functions $U^{i}(r^{1},..., r^{N}), i=1,...,n$ converting any solution of
(\ref{gidra}) to a solution of (\ref{genern}). In hydrodynamics such
solutions
describe nonlinear interaction of $N$ planar waves. Sometimes they  are called $N$-phase solutions.

Clearly, the general solution of (\ref{gidra}) contains $N$ arbitrary functions of one variable.
It turns out that functions $v^{i}, w^{i}$ in the reduction (\ref{gidra}) may contain additional functions of
one variables as functional parameters and the number of these functions is
not greater then  $N$. In \cite{ferhus1} the existence  of hydrodynamic
reductions (\ref{gidra}) locally parameterized by $N$ functions of one variables, where
$N$ is arbitrary, was proposed as a  criterion of integrability for systems
(\ref{genern}). The corresponding  $N$-phase solutions depend on $2 N$
arbitrary functions of one variables.

Usually the integrability of systems (\ref{genern}) is associated with a representation of (\ref{genern}) as  commutativity
conditions for a pair of vector fields \cite{mansan}.
For systems that admit the pseudo\-potential representation \cite{zakh, kr4, odsok1, odsok2} these vector fields are Hamiltonian whereas
for some integrable models the vector fields have more complicated structure. Moreover, for some systems the
vector fields depend on a spectral parameter. Thus it is very difficult  to choose any constructive class of
the vector fields covering all known examples and to propose an universal definition of integrability based on the
commutativity of vector fields. The same problems arise with definitions of integrability given in terms of
dispersionless Lax or zero-curvature representations.

Quite the contrary, the hydrodynamic reduction approach is universal. This means that all integrable models
known by now admit the hydrodynamic reductions. All notions of this approach can be rigorously defined (see  Section 2).
It was demonstrated in \cite{ferhus1} that the existence of hydrodynamic reductions can be algorithmically verified for a
given system  (\ref{genern}) and what is more can be efficiently used for classification of integrable cases.

Families of systems (\ref{gidra}) parameterized by  $N$ functions of one
variables can be described in terms of the so-called systems of Gibbons-Tsarev type
(GT-type systems). The GT-type systems play a crucial role in the approach to integrability based on
the hydrodynamic reductions.

{\bf Definition.} A compatible system of PDEs of the form
$$
\partial_i p_j=f(p_i,p_j,u_1,...,u_n)\,\partial_i u_1, \qquad ~i\ne
j,~i,j=1,...,N,
$$
\begin{equation}\label{gt}
\partial_iu_m=g_m(p_i,u_1,...,u_n)\,\partial_i u_1,\qquad ~m=2,...,n,~i=1,...,N,
\end{equation}
$$
\partial_{i} \partial_{j}u_1=h(p_i,p_j,u_1,...,u_n)\, \partial_i u_1\partial_j u_1,\qquad ~i\ne
j,~i,j=1,...,N
$$
is called $n$-{\it fields GT-type system}.
Here $p_1,...,p_N$, $u_1,...,u_n$ are functions of $r^1,...,r^N,$ $N\geq
3$ and $\partial_i=\frac{\partial}{\partial_{r^i}}$.  Notice that the compatibility conditions give rise to
a system of functional equations for the functions $f,~g_k,~h$ and these equations don't
depend on $N$.

{\bf Example 1} \cite{odsok1}. The system
\begin{equation}\label{u1}
\partial_ip_j=\frac{p_j(p_j-1)}{p_i-p_j}\partial_i u_{1},~~~
\partial_i u_m=\frac{u_m(u_m-1) }{p_i-u_m} \partial_i u_1, \quad
~m=2,...,n,
\end{equation}

\begin{equation}\label{gibtsar2}
\partial_{i} \partial_{j} u_1=\frac{2p_ip_j-p_i-p_j}{(p_i-p_j)^2}\partial_i u_{1}\partial_j u_{1},\qquad
i,j=1,...,N,~~~i\ne j
\end{equation}
is an $n$-field GT-type system for any $n,N$. $\square$ 

The original Gibbons-Tsarev system \cite{gibts} is a degeneration
of (\ref{u1}), (\ref{gibtsar2}). A wide class of integrable systems
(\ref{genern}) related to (\ref{u1}), (\ref{gibtsar2}) is described in \cite{odsok1}. An
elliptic version of this GT-type system and the corresponding integrable 3-dimensional systems were proposed in \cite{odsok2}.

{\bf Definition.} Two GT-type systems are called {\it equivalent} if they are related by a transformation of the form
\begin{equation}\label{gr}
p_i\to\lambda(p_i,u_1,...,u_n),\qquad ~i=1,...,N,
\end{equation}
\begin{equation}\label{gr1}
u_m\to\mu_m(u_1,...,u_n),\qquad ~m=1,...,n.
\end{equation}

{\bf Remark 1.} Our Definitions and formulas (\ref{gt}), (\ref{gr}), (\ref{gr}) admit a coordinates-free interpretation. Let $M$ be a bundle with one-dimensional fiber $E$ and $n$-dimensional base $F$. Then each of $p_i$ is a coordinate on $E$ and $u_1,...,u_n$ are some coordinates on $F$. Thus we obtain a notion of a GT-type structure on $M$. It is likely that there exists a canonical GT-type structure on the natural bundle over the moduli space $M_g$ of genus $g$ algebraic curves. Here $E$ is a curve corresponding to a point in $M_g$. We will not use coordinates-free language in this paper.

For generic GT-type systems the functions $f,h$ have a pole at $p_{i}=p_{j}$.
However, there exist GT-type systems holomorphic at $p_{i}=p_{j}$.

{\bf Example 2.} The system
\begin{equation}\label{triv}
\partial_i p_j=0, \qquad
\partial_i u_m=g_{m}(p_{i}) \partial_i u_1, \qquad \partial_{i} \partial_{j}
u_1=0
\end{equation}
is an $n$-field GT-type system for any $n,N$ and any functions $g_{m}(x)$. Notice that only special choice of
the functions $g_{m}(x)$ gives rise to pairs of compatible semi-Hamiltonian systems (\ref{gidra}). $\square$

In this paper we study systems (\ref{genern}) related to GT-type systems of the form
(\ref{triv}). The main motivation is the following observation.  We examined the 3-dimensional travel
wave reductions for known
examples of integrable $d$-dimensional systems with $d>3$ and found that the GT-type systems corresponding to 
these reductions
are equivalent to (\ref{triv}) with rational functions $g_{k}$. We believe that this observation gives us an algorithm for
constructing of new interesting examples of integrable multi-dimensional systems.

The paper is organized as follows. In Section 2.1 following \cite{ferhus1}, we describe the hydrodynamic reduction method and show that any integrable system (\ref{genern}) is related to a GT-type system.  Section 2.2 is devoted to the return way from GT-type systems to integrable 3-dimensional systems. Moreover, we present all known to us GT-type systems and give a new interpretation of results obtained in \cite{odsok1,odsok2}.

In Section 3 we consider GT-type systems (\ref{triv}) with rational functions $g_{m}(x).$ Using an algorithm described in Section 2.2, we construct the corresponding
families of compatible pairs of hydrodynamic-type 2-dimensional systems, and finally 3-dimensional systems of the form (\ref{genern})
with arbitrary $n,~k$, whose hydrodynamic reductions are given by our 2-dimensional systems. It turns out that all these
3-dimensional systems possess pseudopotential representations with a spectral parameter. In the generic case,
the coefficients of the 3-dimensional systems are expressed in terms of exponents of $u_i$. Degenerations considered in subsection 3.2
involves polynomials in addition to the exponents. In the case of small $n$ and $k$ some of our systems are equivalent to
known dispersionless equations of second order.
In particular, the generic system corresponding to $n=3, ~k=1$ is equivalent to the dispersionless Hirota equation
$$
a_1 Z_x Z_{yt}+a_2 Z_y Z_{xt}+a_3 Z_t Z_{xy}=0, \qquad a_1+a_2+a_3=0.
$$

\vskip.3cm \noindent {\bf Acknowledgments.} Authors thank M.V. Pavlov for fruitful
discussions.   V.S. is grateful to IHES for hospitality and financial support. He was partially supported by
the RFBR grants 08-01-461 and NS 3472.2008.2.

\section{The GT-type systems and integrability}

\subsection{The method of hydrodynamic reductions}

Recall the definitions of the hydrodynamic reduction method and the corresponding criteria of
integrability for 3-dimensional hydrodynamic-type systems \cite{ferhus1}.

{\bf Definition.} An (1+1)-dimensional
hydrodynamic-type system of the form
\begin{equation}\label{gtgen}
r^i_t=\lambda^i(r^1,...,r^N)\, r^i_x, \qquad i=1,...,N,
\end{equation}
is called semi-Hamiltonian if the following relation holds
\begin{equation}   \label{semiham}
\partial_{j}\frac{\partial_{i} \lambda^{m}}{\lambda^{i}-\lambda^{m}}=\partial_{i}\frac{\partial_{j}
\lambda^{m}}{\lambda^{j}-\lambda^{m}}, \qquad  i\ne j\ne m, \qquad
\end{equation}

Semi-Hamiltonian systems have infinitely many symmetries and conservation laws of hydrodynamic type \cite{tsar}.

{\bf Definition.} A hydrodynamic reduction of a system (\ref{genern}) is defined by a pair of compatible semi-Hamiltonian
hydrodynamic-type systems
\begin{equation}\label{gtgen1}
r^i_t=\lambda^i(r^1,...,r^N)\, r^i_x,\qquad r^i_y=\mu^i(r^1,...,r^N)\, r^i_x, \qquad i=1,...,N,
\end{equation}
and by functions $u_1(r^1,...,r^N),...,u_n(r^1,...,r^N)$ such that for each solution of (\ref{gtgen1}) the functions
\begin{equation}\label{red}u_1=u_1(r^1,...,r^N),\,...,\,u_n=u_n(r^1,...,r^N)\end{equation}
satisfy (\ref{genern}).

According to \cite{ferhus1} a system (\ref{genern}) is called {\it integrable} if it
possesses as
many hydro\-dynamic reductions as possible. Namely, substituting (\ref{red}) into (\ref{genern}), eliminating $t$- and $y$-derivatives via
(\ref{gtgen1}), and equating coefficients at $r^l_x$ to zero, we obtain
\begin{equation}   \label{red1}
\sum_{j=1}^n a_{ij}({\bf u})\,\lambda^l\partial_l u_j+\sum_{j=1}^n b_{ij}({\bf
u})\,\mu^l\partial_l u_j+ \sum_{j=1}^n c_{ij}({\bf u})\,\partial_l u_j=0,
\qquad i=1,...,n+k,~l=1,...,N.
\end{equation}
For each fixed $l$ this is a linear overdetermined system for $n$ unknowns $\partial_l u_1,...,\partial_l u_n$, whose coefficients do not depend on $l$.
This linear system must have non-zero solution so all its $n\times n$ minors must be equal to zero.
These minors are polynomials in $\lambda^l,~\mu^l$ independent on $l$. We assume that this system of
polynomial equations is equivalent to one equation
\begin{equation}\label{disp}P(\lambda^l,\mu^l)=0\end{equation}
(otherwise $\lambda^l,~\mu^l$ are fixed and we have not sufficiently many reductions). Equation (\ref{disp})
defines the so-called {\it dispersion algebraic curve}. Let $p$ be a coordinate on this curve. Then (\ref{disp}) is equivalent to equations
$$\lambda^l=F(p_l,u_1,...,u_n),~~~\mu^l=G(p_l,u_1,...,u_n)$$
for some functions $F,~G$. Assume that for generic $p_l$ the linear system (\ref{red1}) has one solution up to
proportionality. Solving this system, we obtain
\begin{equation}\label{gt1}
\partial_iu_m=g_m(p_i,u_1,...,u_n)\,\partial_i u_1,\qquad ~m=2,...,n,~~~i=1,...,N
\end{equation}
for some functions $g_m$. Rewrite (\ref{gtgen1}) in the form
\begin{equation}\label{gtff}
r^i_t=F(p_i,u_1,...,u_n)r^i_x,~~~r^i_y=G(p_i,u_1,...,u_n)r^i_x,\qquad i=1,...,N.
\end{equation}
It is easy to see that the compatibility conditions  for (\ref{gtff}) have the form
\begin{equation}\label{com}\frac{\partial_{i}F(p_j)}{F(p_i)-F(p_j)}=\frac{\partial_{i}G(p_j)}{G(p_i)-G(p_j)}.
\end{equation}
Here we omit arguments $u_1,...,u_n$ in $F,~G$. From (\ref{com}) we can find $\partial_i p_j$ in the form
\begin{equation}   \label{red2}
\partial_i p_j=f(p_i,p_j,u_1,...,u_n)\,\partial_i u_1, \qquad ~i\ne
j,~~~i,j=1,...,N.\end{equation}
Finally, the compatibility conditions $\partial_i\partial_j u_m=\partial_j\partial_i u_m$   give rise to
\begin{equation}   \label{red3}
\partial_{i} \partial_{j}u_1=h(p_i,p_j,u_1,...,u_n)\, \partial_i u_1\partial_j u_1,\qquad ~i\ne
j,~i,j=1,...,N.\end{equation}
Collecting equations (\ref{red1}), (\ref{red2}), (\ref{red3}) together, we obtain a system of the form (\ref{gt}).
  Since we want to have as many reductions as possible,
we assume that this system  is in involution (i.e. fully compatible). In this case the family of hydrodynamic
reductions (\ref{gtff})
locally depends on $N$ functions in one variable.

\subsection{From GT-type systems to integrable models}

In the classification works \cite{ferhus1} the authors start from a class of
systems (\ref{genern}) with fixed small
$n$ and $k$, calculate the corresponding GT-type system and derive
integrability conditions for (\ref{genern}) from the compatibility
conditions for the GT-type system.

In this section we trace the return way and show how to construct wide classes of
integrable systems (\ref{genern}) with arbitrary $n$ and $k$ starting from a
given GT-type system. We also describe our previous results \cite{odsok1, odsok2} from this point of view.

A list of known one-field GT-type systems is given by the following examples.

 {\bf Example 3.} Let $P(x)=a_3 x^3+a_2 x^2+a_1 x+ a_0$. Then
 $$
\displaystyle \partial_{ij}u=\frac{K_2(p_i,p_j) u^2+K_1(p_i,p_j) u+
K_0(p_i,p_j) } {P(u)(p_i-p_j)^2}\,
\partial_i u\partial_j u,$$$$
\displaystyle \partial_ip_j=\frac{P(p_j)(u-p_i)}{P(u)(p_i-p_j)}\partial_i u,~~~i,j=1,...,N,~~~i\ne
j,$$
 where
$$
K_2(p_i,p_j)=2 a_3 (p_i-p_j)^2,$$
$$
K_1(p_i,p_j)=-a_3 (p_i^2p_j+p_ip_j^2)+a_2
(p_i^2+p_j^2-4p_ip_j) - a_1 (p_i+p_j)-2
a_0,$$$$
K_0(p_i,p_j)= 2 a_3 p_i^2 p_j^2+ a_2
(p_i^2p_j+p_ip_j^2)+ a_1 (p_i^2+p_j^2)+a_0
(p_i+p_j)
$$
is an one-field GT-type system.

Using transformations of the form
$$\displaystyle
u\to\frac{a u+b}{c u+d}, \qquad \displaystyle
p_i\to\frac{ap_i+b}{cp_i+d},$$
one can put the
polynomial $P$ to one of the canonical forms: $P(x)=x(x-1)$, $P(x)=x$, or
$P(x)=1$.
$\square$

Note that in the case $P(x)=x(x-1)$ we return to the Example 1 with $n=1$.

{\bf Example 4.}
Let $$
\theta(z, \tau)=\sum_{\alpha\in\Z}(-1)^{\alpha}e^{2\pi i(\alpha z+\frac{\alpha(\alpha-1)}{2}\tau)}, \qquad
\rho(z,\tau)=\frac{\theta_{z}(z,\tau)}{\theta(z,\tau)}.
$$
Then
$$
\partial_{\alpha}p_{\beta}=\frac{1}{2\pi i}\Big(\rho(p_{\alpha}-p_{\beta})-\rho(p_{\alpha})\Big)\partial_{\alpha}
\tau  ,
$$
$$
\partial_{\alpha} \partial_{\beta} \tau=-\frac{1}{\pi i}\rho^{\prime}(p_{\alpha}-p_{\beta})
\partial_{\alpha} \tau \partial_{\beta} \tau,
$$
where $\alpha, \beta=1,...,N, \quad \alpha\ne \beta,$ is an one-field GT-type system.
$\square$

It turns out that if we add the following equations :
$$
\partial_i u_{m}=f(p_i,u_{m},u)\,\partial_i u_{1}, \qquad m=2,...,n
$$
to any one-field GT-type system
$$
\partial_i p_j=f(p_i,p_j,u_{1})\,\partial_i u_{1}, \qquad
\partial_{i} \partial_{j}u_{1}=h(p_i,p_j,u_{1})  \partial_i u_1\partial_j u_1,$$ then the system of PDEs thus obtained is in involution. One can obtain $n$-fields GT-type system for any $n$ in this way.
We call this procedure {\it regular
extension}. For example, in the case of Example 4 the
regular extension is given by
$$
\partial_{\alpha}u_{\beta}=\frac{1}{2\pi i}\Big(\rho(p_{\alpha}-u_{\beta})-\rho(p_{\alpha})\Big)\partial_{\alpha}
\tau,\qquad \beta=1,...,n-1.
$$
The Example 1 is a regular extension of Example 3 with $P(x)=x(x-1)$.
As far as we know, the regular extensions of Examples 3, 4 are the only
GT-type systems appeared in the literature. In this paper we investigate the simplest possible GT-type system
from Example 2 and obtain the corresponding systems of the type (\ref{genern}) which are probably new.

A basic object associated with a given GT-type system is a pair of compatible (1+1)-dimensional hydrodynamic-type systems of the form (\ref{gtff}). One should solve the functional equation  (\ref{com}) in order to find all possible functions $F,~G$. Notice that the derivatives in (\ref{com}) are supposed to be calculated by virtue of the GT-type system.

The existence of non-constant solutions $F,~G$ for the functional equation  (\ref{com}) is an additional
condition, which we impose on the GT-type system. For instance, in the case of
Example 2 the solutions exist not for any functions
$g_{m}.$

The following statement can be proved straightforwardly.

{\bf Proposition 1.} Any one-field GT-type system having a non-constant solution  of the form \linebreak $F(p,u),~G(p,u)$
for the  functional equation (\ref{com}) is equivalent to one described in
Example 3. $\square$

In \cite{odsok1}  we found
the following solutions $F,~G$ for the $n$-field GT-type system in Example 1:

{\bf Proposition 2.}
Fix $s_1,...,s_{n+2}\in\C$. Consider the following compatible overdetermined system of linear PDEs:
$$
\frac{\partial^2 h}{\partial u_j \partial
u_k}=\frac{s_j}{u_j-u_k}\cdot \frac{\partial h}{\partial
u_k}+\frac{s_k}{u_k-u_j}\cdot \frac{\partial h}{\partial
u_j},~~~i,j=1,...,n, \quad j\ne k,
$$
and $$ \begin{array}{c}   \displaystyle \frac{\partial^2
h}{\partial u_j \partial u_j}=-\left(1+\sum_{k=1}^{n+2} s_k\right)
\frac{s_j}{u_j (u_j-1) }\cdot
h+  \frac{s_j}{u_j (u_j-1)} \sum_{k\ne j}^n \frac{u_k
(u_k-1)}{u_k-u_j}\cdot
\frac{\partial h}{\partial u_k}+\\[10mm]
\displaystyle \left(\sum_{k\ne j}^n \frac{s_k}{u_j-u_k}+
\frac{s_j+s_{n+1}}{u_j}+ \frac{s_j+s_{n+2}}{u_j-1}\right)\cdot
\frac{\partial h}{\partial u_j}
\end{array}
$$
It is easy to show that the vector space ${\cal H}$ of all solutions is $n+1$-dimensional.
For any $h\in {\cal H}$ we put
$$
S(h,p)=\sum_{1\leq i\leq
n}u_i(u_i-1)(p-u_1)...\hat{i}...(p-u_n)h_{u_i}+$$$$(1+\sum_{1\leq
i\leq n+2}s_i)(p-u_1)...(p-u_n)h.
$$
Clearly, $S$ is a polynomial of degree $n$ in $p$.

  Let $h_1,h_2,h_3$ be linearly independent elements of ${\cal H}$. Then
\begin{equation}\label{FG}
F=\frac{S(h_1,p)}{S(h_3,p)}, \qquad   G=\frac{S(h_2,p)}{S(h_3,p)}
\end{equation}
satisfy the functional equation (\ref{com}) for reductions.

In the case when the degree of $S$ is less then $n$ the solutions are given
by more complicated determinant formulas (see \cite{odsok1}).
$\square$

For the regular extensions of the Example 4 solutions of the functional equation (\ref{com})
are given by the same formula (\ref{FG}), where
$$
S(h, p)=\sum_{1\leq \alpha\leq
n}\frac{\theta(u_{\alpha})\theta(p-u_{\alpha}-\eta)}{\theta(u_{\alpha}+\eta)
\theta(p-u_{\alpha})}h_{u_{\alpha}}-
(s_1+...+s_n)\frac{\theta^{\prime}(0)\theta(p-\eta)}{\theta(\eta)\theta(p)}h..
$$
Here   $\eta=s_1u_1+...+s_nu_n+r\tau+\eta_0,$ where $s_1,...,s_n,r,\eta_0$ are arbitrary
constants and $h(u_1,...,u_n,\tau)$ is a solution of the following elliptic hypergeometric system:

$$
h_{u_{\alpha}u_{\beta}}=s_{\beta}\Big(\rho(u_{\beta}-u_{\alpha})
+\rho(u_{\alpha}+\eta)-\rho(u_{\beta})
-\rho(\eta)\Big)h_{u_{\alpha}}+$$$$
s_{\alpha}\Big(\rho(u_{\alpha}-u_{\beta})
+\rho(u_{\beta}+\eta)-\rho(u_{\alpha})
-\rho(\eta)\Big)h_{u_{\beta}},
$$

$$
h_{u_{\alpha}u_{\alpha}}=s_{\alpha}\sum_{\beta\ne\alpha}\Big(\rho(u_{\alpha})+
\rho(\eta)-\rho(u_{\alpha}-u_{\beta})-\rho(u_{\beta}+\eta)\Big)h_{u_{\beta}}+
$$$$
\Big(\sum_{\beta\ne\alpha}s_{\beta}\rho(u_{\alpha}-u_{\beta})
+(s_{\alpha}+1)\rho(u_{\alpha}+\eta)+
$$$$
s_{\alpha}\rho(-\eta)+(s_0-s_{\alpha}-1)\rho(u_{\alpha})+2\pi ir
\Big)h_{u_{\alpha}}-
s_0s_{\alpha}(\rho^{\prime}(u_{\alpha})-\rho^{\prime}(\eta))h,
$$

$$ h_{\tau}=\frac{1}{2\pi
i}\sum_{\beta}\Big(\rho(u_{\beta}+\eta)-
\rho(\eta)\Big)h_{u_{\beta}}-\frac{s_0}{2\pi i}\rho^{\prime}(\eta)
h. \qquad \square
$$

Given a GT-type system and a solution $F,~G$ of the functional equation (\ref{com}) for reduction, one can easily
construct an integrable system of the form (\ref{genern}).
Integer $k$ is called the {\it defect} of the system.

{\bf Lemma 1.} Consider the linear space $V$ of functions in $p$ spanned by
$$\{F(p,u_1,...,u_n)g_j(p,u_1,...,u_n),\,
G(p,u_1,...,u_n)g_j(p,u_1,...,u_n),\,
g_j(p,u_1,...,u_n);\quad j=1,...,n\}.$$ Here by definition $g_1=1$. Then the system of the form (\ref{genern}) with reductions (\ref{gtff}) consists of $l$ equations iff $V$
is $(3n-l)$-dimensional.
Moreover, the coefficients of (\ref{genern}) are defined by relations:
$$\sum_{j=1}^n \Big(a_{ij}({\bf u})F(p,u_1,...,u_n)+ b_{ij}({\bf
u})G(p,u_1,...,u_n)+ c_{ij}({\bf u})\Big)g_j(p,u_1,...,u_n)=0,
\qquad i=1,...,n+k.$$

An explicit form of integrable systems (\ref{genern}) corresponding to Examples 3, 4 can be
found in \cite{odsok1, odsok2}.

\section{Weakly nonlinear 3-dimensional systems}

For generic GT-type systems the functions $f,~h$ have poles at $p_{i}=p_{j}$.
However, there exist GT-type systems holomorphic at $p_{i}=p_{j}$.

We call integrable system (\ref{genern}) {\it weakly
nonlinear} if the corresponding GT-type system is holomorphic at $p_{i}=p_{j}$ . 
It is possible to check that if $k=0$, then any 2-dimensional system
describing travel wave solutions ${\bf u}={\bf u}(c_{1} x+c_{2} y+ c_{3} t, \,c_{4} x+c_{5} y+ c_{6} t)$
for weakly nonlinear 3-dimensional system (\ref{genern}) is a weakly nonlinear 2-dimensional system in
the sense of \cite{fer2}.

{\bf Example 5.} Consider the following 3-dimensional system (see \cite{ferhus1}):
\begin{equation} \label{FH}
v_t+a v_x+p v_y+q w_y=0, \qquad w_t+b w_x+r v_y+s w_y=0,
\end{equation}
where
$$
a=w, \quad b=v, \qquad  r=\frac{P(w)}{w-v},  \qquad q=\frac{P(v)}{v-w}, $$
$$\qquad s=\frac{P(v)}{w-v}+\frac{1}{3} P'(v), \qquad p=\frac{P(w)}{v-w}+\frac{1}{3} P'(w).  $$
Here $P$ is an arbitrary polynomial of degree three.

The corresponding GT-type system is given by
$$
\partial_{i} p_{j}=\frac{P(w)}{(w-v) P(v)}\, p_{j}^{2} p_{i}+\left(\frac{1}{w-v}+\frac{P'(v)}{P(v)}
\right)\, p_{j} p_{i}- \left(\frac{1}{v-w}+\frac{P'(w)}{P(w)}
\right)\, p_{j}-\frac{P(v)}{(v-w) P(w)},
$$

$$
\partial_{i} v=p_{i}\, \partial_{i} w,
$$
$$
\partial_{i} \partial_{j} w=\left(\frac{P(w)}{(v-w)P(v)}\, p_{i} p_{j}+\frac{1}{v-w}+\frac{P'(w)}{P(w)}
\right)  \partial_{i} w  \partial_{j} w.
$$
This GT-type system is polynomial in $p_{i}, p_{j}$ and therefore the
corresponding
3-dimensional system is  weakly nonlinear.
It is possible to verify that this GT-type system is equivalent to
$$
\partial_i p_j=0, \qquad
\partial_i u_2=p_{i} \partial_i u_1, \qquad \partial_{i} \partial_{j}
u_1=0.
$$
 $\square$

It was mentioned in \cite{ferhus1} that the system (\ref{FH}) possesses a
hydrodynamic-type Lax representation depending on a spectral parameter. It
turns out that this is a general property of 3-dimensional systems corresponding to
GT-type systems of the form  (\ref{triv}).

{\bf Proposition 3.} Let
$F(p,u_1,...,u_n),~G(p,u_1,...,u_n)$ be a solution of the functional equation (\ref{com}) for a GT-type system (\ref{triv}). Then the corresponding 3-dimensional system admits the Lax representation
$$\psi_{t}=F(\xi,u_1,...,u_n)\, \psi_x,~~~~~~~\psi_{y}=G(\xi,u_1,...,u_n)\, \psi_x,$$
where $\xi$ is a spectral parameter. $\square$

\subsection{Generic case}

Using our observation that the GT-type system from Example 5 is equivalent to  (\ref{triv}) with rational functions
$g_{m}$, we generalize Example 5 to the case of arbitrary $n$ and $k$.

Consider the  $(n+1)$-field GT-type system (\ref{triv}) with
$g_{m}=M_{m}/M,$ where $M, M_{1},...,M_{n+1}$ are generic polynomials of degree $n$.
Suppose that $M$ has pairwise distinct roots $\lambda_0,
\lambda_1,...,\lambda_n$. Then up to equivalence the GT-type system can be
written as
\begin{equation}\label{gibtsar1}
\partial_{i} p_{j}=0,\qquad  \partial_{i}u_m=\frac{\lambda_m-\lambda_0}{p_i-\lambda_m}\partial_{i}w, \qquad
\partial_{i}\partial_{j}w=0
\end{equation}
with fields denoted by $u_1,...,u_n,~w$.

Let $H_n$ be the linear space of functions in $u_1,...,u_n$ spanned by $1,~e^{u_1},...,e^{u_n}$.
For any function $g=a_0+a_1e^{u_1}+...+a_ne^{u_n}\in H_n$ we put
$$S_n(g,p)=\frac{a_0}{p-\lambda_0}+
\sum_{i=1}^n \frac{a_i e^{u_i}}{p-\lambda_i}.$$

For $k\in\N$ such that $0<k<n-1$ we fix functions $h_1,...,h_k\in H_n$, where $ h_i=b_{i,0}+b_{i,1}e^{u_1}+...+b_{i,n}e^{u_n},$
and define
\begin{equation}\label{polgen}
S_{n,k}(g,p)=\det\left(\begin{array}{ccccc}S_n(g,p)&S_n(h_1,p)&...&S_n(h_k,p)
\\g&h_1&...&h_k\\a_{n-k+2}&b_{1,n-k+2}&...&b_{k,n-k+2}
\\.........&...&...&.........\\a_n&b_{1,n}&...&b_{k,n}
\end{array}\right)\,.
\end{equation}
By definition, $S_{n,0}(g,p)=S_{n}(g,p)$.

{\bf Proposition 4.} Let $g_1,~g_2,~g_3$ be linearly independent elements of
$H_n$. Then for any  $0\le k<n-1$ the functions
\begin{equation}\label{gidragencom}
F=\frac{S_{n,k}(g_1,p_i)}{S_{n,k}(g_3,p_i)}, \qquad
G=\frac{S_{n,k}(g_2,p_i)}{S_{n,k}(g_3,p_i)}
\end{equation} satisfy the functional equation (\ref{com}) for hydrodynamic reductions.
$\square$

To find an explicit form of the corresponding 3-dimensional systems we
note that
$$\sum_{i=1}^n (A_iu_{i,t_1}+B_iu_{i,t_2}+C_iu_{i,x})=0$$
is an equation from the  3-dimensional system iff
$$\sum_{i=1}^n\frac{\lambda_i-\lambda_0}{p-\lambda_i}\Big(A_iS_{n,k}(g_1,p)+B_iS_{n,k}(g_2,p)+C_iS_{n,k}(g_3,p)\Big)=0$$
as function in $p$.
Let $g_i=a_{i,0}+a_{i,1}e^{u_1}+...+a_{i,n}e^{u_n},~i=1,2,3$.

If $k=0$, then the corresponding 3-dimensional system reads as follows:

$$\sum_{1\leq j\leq n, j\ne i}(a_{2,i}a_{3,j}-a_{2,j}a_{3,i})e^{u_j}\frac{u_{i,t_1}-u_{j,t_1}}{\lambda_i-\lambda_j}+(a_{2,i}a_{3,0}-a_{3,i}a_{2,0})\frac{u_{i,t_1}}{\lambda_i-\lambda_0}+$$
\begin{equation}\label{3d0}\sum_{1\leq j\leq n, j\ne i}(a_{3,i}a_{1,j}-a_{3,j}a_{1,i})e^{u_j}\frac{u_{i,t_2}-u_{j,t_2}}{\lambda_i-\lambda_j}+(a_{3,i}a_{1,0}-a_{1,i}a_{3,0})\frac{u_{i,t_2}}{\lambda_i-\lambda_0}+\end{equation}
$$\sum_{1\leq j\leq n, j\ne i}(a_{1,i}a_{2,j}-a_{1,j}a_{2,i})e^{u_j}\frac{u_{i,x}-u_{j,x}}{\lambda_i-\lambda_j}+
(a_{1,i}a_{2,0}-a_{2,i}a_{1,0})\frac{u_{i,x}}{\lambda_i-\lambda_0}=0,$$
where $i=1,...,n$.

If $k>0$, then the corresponding 3-dimensional system reads as follows:

$$
\sum_{1\leq j\leq n-k+1,j\ne
i}\Big(\Delta_i(g_2)\Delta_j(g_3)-\Delta_j(g_2)\Delta_i(g_3)\Big)e^{u_j}\frac{u_{i,t_1}-u_{j,t_1}}{\lambda_i-\lambda_j}+\Big(\Delta_i(g_2)\Delta_0(g_3)-\Delta_0(g_2)\Delta_i(g_3)\Big)\frac{u_{i,t_1}}{\lambda_i-\lambda_0}+
$$
$$
\sum_{1\leq j\leq n-k+1,j\ne
i}\Big(\Delta_i(g_3)\Delta_j(g_1)-\Delta_j(g_3)\Delta_i(g_1)\Big)e^{u_j}\frac{u_{i,t_2}-u_{j,t_2}}{\lambda_i-\lambda_j}+\Big(\Delta_i(g_3)\Delta_0(g_1)-\Delta_0(g_3)\Delta_i(g_1)\Big)\frac{u_{i,t_2}}{\lambda_i-\lambda_0}+
$$
$$
\sum_{1\leq j\leq n-k+1,j\ne
i}\Big(\Delta_i(g_1)\Delta_j(g_2)-\Delta_j(g_1)\Delta_i(g_2)\Big)e^{u_j}\frac{u_{i,x}-u_{j,x}}{\lambda_i-\lambda_j}
+\Big(\Delta_i(g_1)\Delta_0(g_2)-\Delta_0(g_1)\Delta_i(g_2)\Big)\frac{u_{i,x}}{\lambda_i-\lambda_0}=0,
$$
where $i=1,...,n-k+1$ and
$$
\sum_{j=1}^{n-k+1}e^{u_j}\Delta_j(g_r)u_{j,t_s}=\sum_{j=1}^{n-k+1}e^{u_j}\Delta_j(g_s)u_{j,t_r},
$$
$$
\sum_{j=1}^{n-k+1}\Delta_j(g_r)e^{u_j}\frac{u_{i,t_s}-u_{j,t_s}}{\lambda_i-\lambda_j}+
\Delta_0(g_r)\frac{u_{i,t_s}}{\lambda_i-\lambda_0}=
\sum_{j=1}^{n-k+1}\Delta_j(g_s)e^{u_j}\frac{u_{i,t_r}-u_{j,t_r}}{\lambda_i-\lambda_j}+
\Delta_0(g_s)\frac{u_{i,t_r}}{\lambda_i-\lambda_0},
$$
where $i=n-k+2,...,n$. Here $r,s=1,2,3$, $t_3=x$ and
$$\Delta_j(g)=\det\left(\begin{array}{cccc}g&h_1&...&h_k\\a_j&b_{1,j}&...&b_{k,j}
\\a_{n-k+2}&b_{1,n-k+2}&...&b_{k,n-k+2}
\\.........&...&...&.........\\a_n&b_{1,n}&...&b_{k,n}
\end{array}\right)\,$$
for $j=1,...,n$, $g=a_0+a_1e^{u_1}+...+a_ne^{u_n}$, $h_1=b_{1,0}+b_{1,1}e^{u_1}+...+b_{1,n}e^{u_n}$,...,$h_k=b_{k,0}+b_{k,1}e^{u_1}+...+b_{k,n}e^{u_n}$. Note that this equations are linearly dependent and the system is equivalent to a system with $n+k$ linearly independent  equations.

{\bf Proposition 5.} If $k=0$, then the corresponding system (\ref{3d0}) possesses the following $n+1$ conservation laws of hydrodynamic type:
$$
\left|\begin{array}{cc}S^{reg}_n(g_2,\lambda_i)&S^{reg}_n(g_3,\lambda_i)\\a_{2,i}e^{-u_i}&a_{3,i}e^{-u_i}\end{array}\right|_{t_1}+\left|\begin{array}{cc}S^{reg}_n(g_3,\lambda_i)&S^{reg}_n(g_1,\lambda_i)\\a_{3,i}e^{-u_i}&a_{1,i}e^{-u_i}\end{array}\right|_{t_2}+\left|\begin{array}{cc}S^{reg}_n(g_1,\lambda_i)&S^{reg}_n(g_2,\lambda_i)\\a_{1,i}e^{-u_i}&a_{2,i}e^{-u_i}\end{array}\right|_x=0,
$$
where $i=1,...,n$ and 
$$
\left|\begin{array}{cc}S^{reg}_n(g_2,\lambda_0)&S^{reg}_n(g_3,\lambda_0)\\a_{2,0}&a_{3,0}\end{array}\right|_{t_1}+\left|\begin{array}{cc}S^{reg}_n(g_3,\lambda_0)&S^{reg}_n(g_1,\lambda_0)\\a_{3,0}&a_{1,0}\end{array}\right|_{t_2}+\left|\begin{array}{cc}S^{reg}_n(g_1,\lambda_0)&S^{reg}_n(g_2,\lambda_0)\\a_{1,0}&a_{2,0}\end{array}\right|_x=0.
$$
Here 
$$S^{reg}_n(g,\lambda_i)=\Big(S_n(g,p)-\frac{a_ie^{u_i}}{p-\lambda_i}\Big)\Big|_{p=\lambda_i}$$
and
$$S^{reg}_n(g,\lambda_0)=\Big(S_n(g,p)-\frac{a_0}{p-\lambda_0}\Big)\Big|_{p=\lambda_0}.$$

{\bf Proposition 6.} If $k>0$, then the corresponding system possesses the following $3k$ conservation laws of hydrodynamic type:
\begin{equation}   \label{conlaw}
\left(\frac{\Delta(g_r,h_1,...\hat{i}...h_k)}{\Delta(h_1,...,h_k)}\right)_{t_s}=
\left(\frac{\Delta(g_s,h_1,...\hat{i}...h_k)}{\Delta(h_1,...,h_k)}\right)_{t_r},
\end{equation}
where $i=1,...,k$, $r,s=1,2,3$, $t_3=x$ and
$$\Delta(f_1,...,f_k)=\det\left(\begin{array}{ccc}f_1&...&f_k\\f_{1,u_{n-k+2}}&...&f_{k,u_{n-k+2}}
\\.........&...&.........\\ f_{1,u_n}&...&f_{k,u_n}
\end{array}\right).$$

{\bf Remark 2.} It is likely that for $k>0$ the corresponding 3-dimensional system possesses additional $n+1$ conservation laws of hydrodynamic type.

{\bf Remark 3.} Proposition 6 allows us to define functions $z_1,...,z_k$ such that
\begin{equation}   \label{z}
\frac{\Delta(g_r,h_1,...\hat{i}...h_k)}{\Delta(h_1,...,h_k)}=z_{i,t_r}
\end{equation}
for all $i=1,...,k$ and $r=0,1,3$. See \cite{odsok1} or \cite{odsok2} for further discussion.

\subsection{Degenerations}

Our constructions of GT-type systems and functions $S_{n,k}$ are valid in the case of pairwise distinct roots $\lambda_0,...,\lambda_n$.
In this section we study degenerations of the GT-type systems described in Section 3.1.

Define polynomials $P_i(u_1,u_2,...)$ as coefficients of the following Taylor expansion
$$
\exp{(\varepsilon u_1+\varepsilon^2 u_2+...)}=1+P_1 \varepsilon+P_2 \varepsilon^2+\cdots.
$$
In particular,
$$
P_1=u_1, \quad P_2=u_2+\frac{1}{2} u_1^2, \quad P_3=u_3+u_1 u_2+\frac{1}{6} u_1^3.
$$
Denote the partial sums $1+\sum_{i=1}^k P_i \varepsilon^i$ by $Q_k(\varepsilon)$. By definition, $P_0=Q_0(\varepsilon)=1.$

{\bf Degeneration 1}. This degeneration corresponds to the case $\lambda_0\ne\lambda_1=...=\lambda_n$.
Consider the following $(n+1)$-field GT-type system
with fields $u_1,...,u_n,~w$:
\begin{equation}\label{gibtsar11}
\partial_{i} p_{j}=0,~~~ \partial_{i}u_m=\left(\frac{\lambda_1-\lambda_0}{(p_i-\lambda_1)^m}+
\frac{1}{(p_i-\lambda_1)^{m+1}} \right)\partial_{i}w,~~~\partial_{i}\partial_{j}w=0.
\end{equation}

For any vector $(a_0, a_1, ... , a_n)$ define
$$
g=a_0+e^{u_1} \sum_{i=1}^n a_i P_{i-1}
$$
and
$$S_n(g,p)=\frac{a_0}{p-\lambda_0}+
e^{u_1} \sum_{i=1}^n \frac{a_i Q_{i-1}(p-\lambda_1)}{(p-\lambda_1)^{i}}.$$

{\bf Degeneration 2}. This degeneration corresponds to the case $\lambda_0=\lambda_1=...=\lambda_n$.
Consider the following $(n+1)$-field GT-type system:
\begin{equation}\label{gibtsar12}
\partial_{i} p_{j}=0,\qquad \partial_{i}u_m= \frac{1}{(p_i-\lambda_0)^m} \partial_{i}w,\qquad \partial_{i}\partial_{j}w=0.
\end{equation}

For any vector $(a_0, a_1, ... , a_n)$ define
$$
g=  \sum_{i=0}^n a_i P_{i}.
$$
and
$$S_n(g,p)= \sum_{i=0}^n \frac{a_i Q_i(p-\lambda_0)}{(p-\lambda_0)^{i+1}}.$$

Combining these degenerations, one obtains the general case.  Let $\lambda_0,...,\lambda_l$ be
pairwise distinct roots of multiplicities $n_0+1,n_1...,n_l$ correspondingly. Note that $n_0+...+n_l=n$.
Consider the following $(n+1)$-field system
with fields
$u_{0,1},...,u_{0,n_0},u_{1,1},...,u_{1,n_1},...,u_{l,1},...,u_{l,n_l}~w$:
\begin{equation}\label{gibtsar13}
\partial_{i} p_{j}=0,~~~\partial_{i}u_{0,m}= \frac{1}{(p_i-\lambda_0)^m} \partial_{i}w,~~~
\partial_{i}u_{s,m}=\left(\frac{\lambda_s-\lambda_0}{(p_i-\lambda_s)^m}+
\frac{1}{(p_i-\lambda_s)^{m+1}} \right)\partial_{i}w,~~~\partial_{i}\partial_{j}w=0.
\end{equation}

For any vector $(a_{0,0}, a_{0,1}, ... , a_{0,n_0},a_{1,1},...,a_{1,n_1},...,a_{l,n_l})$ define
$$g= \sum_{i=0}^{n_0} a_{0,i} P_i+\sum_{s=1}^l\sum_{i=1}^{n_s} a_{s,i}e^{u_{s,i}}  P_{i-1}$$ and
$$S_n(g,p)= \sum_{i=0}^{n_0} \frac{a_{0,i} Q_i(p-\lambda_0)}{(p-\lambda_0)^{i+1}}+\sum_{s=1}^l\sum_{i=1}^{n_s} \frac{a_{s,i} e^{u_{s,i}} Q_{i-1}(p-\lambda_s)}{(p-\lambda_s)^{i}}.$$

Let $k>0$. Fix vectors $(b_{0,0,j}, b_{0,1,j}, ... , b_{0,n_0,j},b_{1,1,j},...,b_{1,n_1,j},...,b_{l,n_l,j}),~j=1,...,k$. Let
$$g= \sum_{i=0}^{n_0} a_{0,i} P_i+\sum_{s=1}^l\sum_{i=1}^{n_s} a_{s,i}e^{u_{s,i}}  P_{i-1}$$
and similarly
$$h_j= \sum_{i=0}^{n_0} b_{0,i,j} P_i+\sum_{s=1}^l\sum_{i=1}^{n_s} b_{s,i,j}e^{u_{s,i}}  P_{i-1},~j=1,...,k.$$
Define $S_{n,k}(g,p)$ by
\begin{equation}\label{polgendeg}
S_{n,k}(g,p)=\det\left(\begin{array}{ccccc}S_n(g,p)&S_n(h_1,p)&...&S_n(h_k,p)
\\g&h_1&...&h_k\\g_{{\bf v}_{n-k+2}}&h_{1,{\bf v}_{n-k+2}}&...&h_{k,{\bf v}_{n-k+2}}
\\.........&...&...&.........\\g_{{\bf v}_n}&h_{1,{\bf v}_n}&...&h_{k,{\bf v}_n}
\end{array}\right)
\end{equation}
where $({\bf v}_1,...,{\bf
v}_n)=(u_{0,1},...,u_{0,n_0},u_{1,1},...,u_{1,n_1},...,u_{l,1},...,u_{l,n_l})$.
Then Proposition 4 holds. The explicit form of the corresponding 3-dimensional systems can be 
calculated using the general recipe.

\section{Summary and discussion: toward a classification of integrable 3-dimensional hydrodynamic-type systems}

We summarize our previous remarks as the following project of classification of integrable 
3-dimensional hydrodynamic-type systems:

{\bf 1.} Classify (up to equivalence (\ref{gr}), (\ref{gr1})) all GT-type systems (\ref{gt}) possessing a non-trivial solution of the functional equation (\ref{com}).

{\bf 2.} For each such GT-type system classify all solutions of the functional equation (\ref{com}).

{\bf 3.} For each GT-type system and solution of (\ref{com}) construct the corresponding 3-dimensional system (see Lemma 1).

This approach has the following advantages:

{\bf a.} It turns out that GT-type systems are universal: for each GT-type system there exist several families of the corresponding 3-dimensional systems depending on essential parameters. Moreover, it is likely that for each genus $g=0,1,...$ 
there exists an unique GT-type system (with one field for $g=0,~1$ and with $3g-3$ fields for $g>1$) such that any generic 3-dimensional system with the dispersion curve of genus $g$ corresponds to this GT-type system or its regular extension.

{\bf b.} Integrable systems (\ref{genern}) are defined up to arbitrary point transformations. Since the equivalence problem for systems (\ref{genern}) is highly non-trivial, it is not easy to find the simplest coordinates in 
which a given 3-dimensional system has the simplest form. As the rule, the simplest coordinates for the GT-type system are at the same time the most natural coordinates for the corresponding system (\ref{genern}). For example, each of the variables $p_i, u$ from the Example 3 admits natural interpretation as a coordinate on $\C P^1$. Similarly, each of the variables $p_i$ from the Example 4 can be naturally interpreted as a coordinate on an elliptic curve and $\tau$ as a modular parameter of this curve. We expect that for $g>1$ there exists a canonical GT-type system with $3g-3$ fields being coordinates on the moduli space of genus $g$ curves. This GT-type system should have (an analog of) regular extensions with additional fields being points on this curve. One could find this GT-type system by study of hydrodynamic reductions for 3-dimensional systems from \cite{kr4}. On the other hand, taking into account the results related to $g=0,1$  we expect that there are much more 3-dimensional systems corresponding to the canonical GT-type system, than it was found in \cite{kr4}. 

The main disadvantage of the approach outlined above is that the realization of the item {\bf 1} is very 
hard for $n>1$. We hope to address this classification problem later. In this paper we consider the simplest possible GT-type system of the form (\ref{triv}) and demonstrate that even in this case there exists a rich family of interesting integrable 3-dimensional systems associated with (\ref{triv}). The coefficients of these systems are quasi-polynomials. 
Note that slightly more complicated GT-type systems from 
Examples 3 and 4 require generalized hypergeometric functions for description of the corresponding 3-dimensional systems.

\end{document}